\documentclass[aps,pra,reprint,superscriptaddress,10pt,showpacs,a4paper]{revtex4-1}
\usepackage{amsmath,amssymb,amsfonts}
\usepackage{mathtools}
\usepackage{graphicx}
\usepackage{txfonts}
\usepackage{color, xcolor}

\usepackage[unicode,breaklinks]{hyperref}
\hypersetup{
    unicode=true,
    a4paper=true,
    plainpages=false,
    pdftitle={Numerical method for the projected Gross--Pitaevskii equation in an infinite rotating 2D Bose gas},
    pdfauthor={R. Doran and T. P. Billam},
    pdfsubject={Numerical method for the projected Gross--Pitaevskii equation in an infinite rotating 2D Bose gas},
    colorlinks=true,
    linkcolor=blue,
    citecolor=blue,
    filecolor=black,
    urlcolor=blue
}
\usepackage{tikz}

\newcommand{\eqnreft}[1]{{Eq.~(\ref{#1})}}

\newcommand{\Arg}{\text{Arg}}

\begin{document}
\title{Numerical method for the projected Gross--Pitaevskii equation in an infinite rotating 2D Bose gas}

\author{R. Doran}
\affiliation{Joint Quantum Centre (JQC) Durham--Newcastle, Department of Mathematics, Statistics and Physics, Newcastle University, Newcastle upon Tyne, NE1 7RU, UK}
\author{T. P. Billam}
\affiliation{Joint Quantum Centre (JQC) Durham--Newcastle, Department of Mathematics, Statistics and Physics, Newcastle University, Newcastle upon Tyne, NE1 7RU, UK}

\date{\today}

\begin{abstract}
We present a method for evolving the projected Gross-Pitaevskii equation in an infinite rotating Bose-Einstein condensate, the ground state of which is a vortex lattice. We use quasi-periodic boundary conditions to investigate the behaviour of the bulk superfluid in this system, in the absence of boundaries and edge effects. We also give the Landau gauge expression for the phase of a BEC subjected to these boundary conditions. Our spectral representation uses the eigenfunctions of the one-body Hamiltonian as basis functions. Since there is no known exact quadrature rule for these basis functions we approximately implement the projection associated with the energy cut-off, but show that by choosing a suitably fine spatial grid the resulting error can be made negligible. We show how the convergence of this model is affected by simulation parameters such as the size of the spatial grid and the number of Landau levels. Adding dissipation, we use our method to find the lattice ground state for $N$ vortices. We can then perturb the ground-state, in order to investigate the melting of the lattice.   
\end{abstract}

\maketitle

\section{Introduction}
\label{sec_introduction}
One of the most striking properties of Bose-Einstein Condensates (BECs) is the effect of forcing them to rotate \cite{Donnelly}. Unlike the solid body rotation of a normal fluid, when a BEC rotates an array of quantised vortices is formed \cite{abo2002formation}. 
Since these quantized vortices were observed experimentally in a BEC \cite{Matthews1999}, they have been a widely studied quantum phenomenon \cite{Bewley08b,Engels2004,Freilich10a,Haljan2001,Henn09a,ketterle_review, Engels2003, Schweikhard04}. Systems with a large number of vortices have been revealed to display a rich selection of dynamics such as the dipole interactions of vortices with opposite charges \cite{Aioi2011}, the mechanisms of vortex lattice formations \cite{ Tsubota2002, Kasamatsu2003, Lobo2004, Bradley08a, Parker2006rotating, Sinha2001, Lobo2004}, and vortex turbulence \cite{Wright08a, Mizushima2004, ParkerPRL05}.

The most common theoretical description of these systems is the zero-temperature, mean-field Gross--Pitaevskii equation (GPE). A wide range of numerical methods have been applied to solving this equation, both with and without rotation. Examples include Crank—Nicolson schemes \cite{Tsubota2002, Kasamatsu2003, Aftalion2003, Aftalion2004, Murangananandam2009, Kumar2019}, backwards Euler finite difference schemes \cite{Bao2005, Bao2004, Bao2006, Antoine2014}, and Sobolov Gradient Methods for a rotating condensate \cite{Garcai2001_PRA, Garcia2001_SIAM, Danaila2010, Vergez2016}. A range of (pseudo-) spectral methods have also been used with, for example, Fourier~\cite{javanainen_ruostekoski_jpa_2006}, Chebyshev~\cite{Jeng2013}, and Hermite~\cite{Dion2003} basis functions.

The \textit{projected} Gross--Pitaevskii equation (PGPE) \cite{Davis2001b} is a classical field equation for simulating a weakly interacting Bose gas at finite temperatures. The PGPE is a microcanonical equation of motion, and the atom number and total energy are conserved quantities. Its crucial feature, beyond the ordinary, non-projected Gross--Pitaevskii equation, is precise implementation of an energy cutoff in the basis of non-interacting single-particle modes. When working at finite temperature, this allows one to set the cutoff such that all included modes have occupation $\gtrsim 1$; in this regime quantum fluctuations are relatively small and the classical field description is accurate. The importance of implementing the projection in the correct non-interacting single-particle basis has been demonstrated \cite{bradley_properties_2005}.  Ideally, the numerical projection operation used to evolve the equation should be numerically exact, necessitating a (pseudo-)spectral approach the implementation. Consequently, although it imparts the ability to describe finite temperature gases, one can also view the PGPE as a systematically dealiased pseudospectral method for the ordinary GPE \cite{boyd}; the wavefunction is described as an expansion over a finite number of basis functions and evolved precisely according to the equation of motion. Taking this view, using a well-defined energy cutoff in the single-particle basis remains advantageous. The PGPE sits within a broader range of techniques known as the c-field methodology \cite{Davis2001a, Davis2001b, Davis2002, gardiner_davis_jpb_2003, Blakie05a, Bradley08a, c-field_review}.

The dynamics of  Rotating 2D Bose gases have been previously studied with the PGPE \cite{ Wright08a,Wright10} in finite, harmonically-trapped system using a Laguerre-Gaussian basis. However, in simulations where the condensate has an edge, vortices nucleate at the interface between the condensate and the thermal cloud. These vortices do not penetrate the main bulk of the condensate, rather they remain at the edge of the condensate for considerable time \cite{Wright08a}. Between these edge effects, and the tendency of the trapping potential to distort any resulting vortex lattice \cite{Sheehy_inhomogenous,Fetter_LLL_2007}, it difficult to conduct a PGPE simulation of sufficient size to isolate the bulk properties of the system \cite{Wright08a}.

In order to concentrate on the bulk of the system and avoid boundary effects --- in a similar way as would be achieved using periodic boundary conditions in the non-rotating case ---
previous works on rotating 2D systems have used quasi-periodic boundary conditions to simulate a representative cell of an infinite rotating system. Physically this corresponds to a harmonically trapped gas, rotating rapidly enough that the effective harmonic trapping vanishes. Under such rapid rotation, if the number of vortices in the Bose gas approaches the number of atoms, the gas enters a fractional quantum Hall regime and the classical field approach of the PGPE breaks down. Here, we consider the alternative regime where the number of vortices remains small compared to the number of atoms and the PGPE remains valid. This regime itself breaks down in to two cases. In the first case, at low temperatures, and low interaction energies such that the typical spacing between vortices in a ground state lattice is comparable to the vortex core size, the Lowest Landau Level approximation can be used to determine the ground state of a system with good accuracy \cite{Butts1999, TL-Ho2001, aftalion2005vortex}. Such an approximation has been used extensively to study vortex lattices \cite{Mueller2002, cooper2004vortex, watanabe2004landau, aftalion2005vortex, Sonin05, aftalion2006vortex, Fetter_LLL_2007, Matveenko2009}, however, it is necessarily limited to the lowest energy states of the system. In the second case, at higher temperatures, with nonequilibrium dynamics, or simply with higher interaction energies such that the typical spacing between vortices in a ground state lattice is much greater than the vortex core size, higher-energy states than the lowest Landau levels must be included. Calculations for an infinite ground state vortex lattice in this case are described in Ref.~\cite{Cozzini2006}, and for dipolar gases in Ref.~\cite{komineas_vortex_2007}. The time-dependent GPE has been implemented in this case with quasi-periodic boundary conditions in Refs. \cite{mingarelli2016simulating, wood2019quasiperiodic}, by using magnetic Fourier transforms and finite difference methods in the symmetric gauge respectively.  However, these methods do not operate directly in a basis of single-particle eigenstates, making it difficult to implement the projection operation needed for the PGPE.

\begin{figure}[t]
	\centering
	\includegraphics[width=\linewidth]{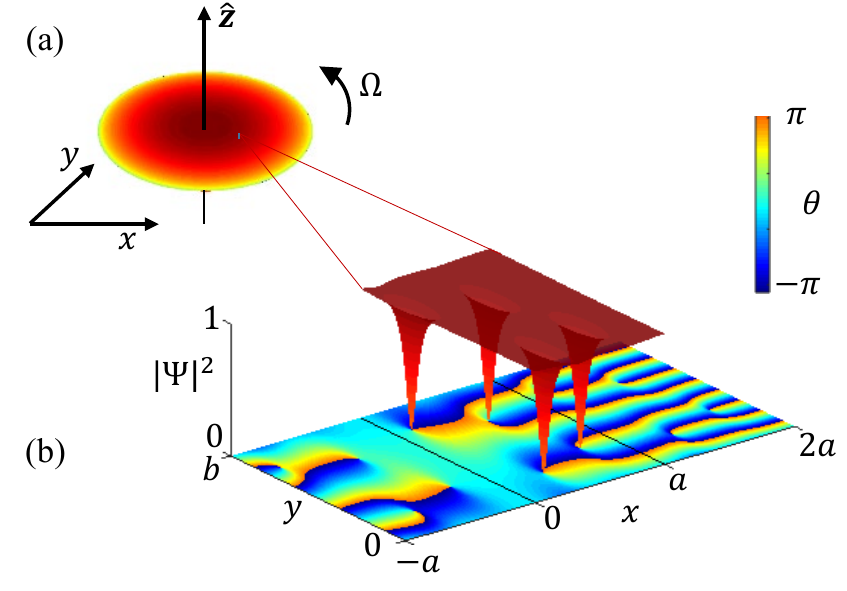}
	\caption{A sketch of the system: (a) A large, oblate, harmonically trapped ($\omega_x = \omega_y = \omega_\perp \ll \omega_z = \omega_\parallel$) condensate rotating with angular frequency $\Omega$. (b) In the centrifugal limit ($\Omega \rightarrow \omega_\perp$) a small cell in the bulk of the now-infinite condensate can be approximated using the Landau gauge with quasi-periodic (twisted) boundary conditions. The height of the surface represents the density of the wavefunction, while the colour represents the phase of the superfluid.}
	\label{fig:systemsketch}
\end{figure}

In this work we present a numerical method for simulating the PGPE in an infinite rotating 2D Bose gas. Our method operates in the Landau gauge, using the correct single-particle basis under quasi-periodic boundary conditions for a representative cell of the system (Fig.~\ref{fig:systemsketch}). By establishing a method to integrate the PGPE for such a rotating system, we open the door to study finite-temperature, non-equilibrium dynamics of rotating systems in the bulk, free of edge effects.

  The remainder of this paper is structured as follows: In Sec.~\ref{sec_PGPE} we introduce the equation of motion which governs a harmonically trapped Bose gas rotating at the centrifugal limit, as well as introducing the quasi--periodic boundary conditions which we use throughout the paper. In Sec.~\ref{sec_num-method} we introduce the PGPE for a rotating system; we also quantify the error which is due to the projection. In Sec.~\ref{sec_vortex-ansatz} we detail how our method allows one to choose an arbitrary array of vortices as an initial condition for the PGPE. This requires us to find the Landau gauge expression for the phase of $N$ vortices.  Sec.~\ref{sec_convergence-and-testing} contains the main results of the paper: we consider how the PGPE evolution performs for varying simulation parameters, as well as looking at how our method can be used to find the ground state of a given system. We then investigate how stable this ground state is. In Sec.~\ref{sec_application} we perturb the ground state of the system, in order to investigate how the lattice responds to melting.


\section{Rotating Projected Gross-Pitaevskii Equation}
\label{sec_PGPE}
\subsection{Single-Particle Hamiltonian}
\label{subsec_EOM}

In the rotating frame, the Hamiltonian for a particle of mass $m$ rotating with angular momentum $\boldsymbol{\Omega}$ is \cite{Landau&Lifshitz}
\begin{equation} 
H_\Omega = \frac{| \boldsymbol{p} |^2 }{2m} + \frac{1}{2} m \omega_\perp^2 \left( x^2 + y^2 \right) + \frac{1}{2} m \omega_\parallel^2 z^2 - \boldsymbol{\Omega \cdot r} \times \boldsymbol{p} ,
\end{equation}
where $\omega_\parallel$ and $\omega_\perp$ are the trapping frequencies in the $z$ and the radial directions, respectively. Throughout this paper, we will not worry about non-uniform rotation, disturbance to the density of the fluid, or any other affects which would be a direct result of the mechanism used to spin the gas. On choosing the $z$ axis to be the axis of rotation, $\boldsymbol{\Omega}=\Omega\boldsymbol{\hat{z}},$ the Hamiltonian may be written as  \cite{cooper2001quantum, cooper2008rapidly, komineas2012vortex} 
\begin{equation}
H_\Omega = \frac{ \left( \boldsymbol{p} - m \boldsymbol{\Omega} \times \boldsymbol{r} \right)^2}{2m} + \frac{1}{2} m \left( \omega_\perp^2 - \Omega^2 \right) \left( x^2 + y^2 \right) + \frac{1}{2}m \omega_\parallel^2 z^2 .
\label{eqn_factorised-hamiltonian}
\end{equation}
 In the middle term of Eqn.~\eqref{eqn_factorised-hamiltonian} we see that the frequency of rotation $\Omega$ reduces the radial trapping frequency. We  set $\Omega=\omega_\perp,$ which is defined in Ref.  \cite{cooper2008rapidly} as the centrifugal limit \footnote{Experimentally it is possible to achieve $\Omega = 0.99 \omega_\perp,$ see for example \cite{Engels2004, Schweikhard04}}. This yields the Hamiltonian
\begin{equation}
H_\Omega = \frac{\left( \boldsymbol{p} - \boldsymbol{A}\right)^2}{2m} + \frac{1}{2} m \omega_\parallel^2 z^2,
\label{eqn_reduced-hamiltonian}
\end{equation}
where the quantity $H_\Omega \Psi$ is invariant under the transformation
\begin{equation}
    \boldsymbol{A} \to \boldsymbol{A} + \nabla \lambda, \qquad \Psi \to \exp \left( \frac{i}{\hbar} \lambda \right) \ \Psi,
    \label{eqn:gauge_symmetry}
\end{equation}
for a given $\lambda$, a function of $x$ and $y$. Hence we have the gauge freedom to choose any $\boldsymbol{A}$ such that $\nabla \times \boldsymbol{A} = 2m \Omega \boldsymbol{\hat{z}}$.
Eqn.~\eqref{eqn_factorised-hamiltonian} is implicitly in the symmetric gauge, which is logical outside the centrifugal limit, as the single particle basis functions are the associated Laguerre polynomials \cite{Fock28}.

The trapping of a BEC gives rise to several boundary phenomena, including the short lived nucleation and annihilation of vortices which do not penetrate the bulk of the fluid \cite{Wright08a}. At the centrifugal limit, it is advantageous to use the  Landau gauge,
\begin{equation}
\boldsymbol{A} = \left( \begin{matrix} 0 \\ 2m \Omega x \end{matrix} \right),
\label{eqn_Landaugauge}
\end{equation}
as the single particle basis functions with quasi-periodic boundary conditions can be found. This will enable us to study the bulk of the Bose gas using the PGPE, without worrying about edge effects.


\subsection{The GPE in Dimensionless Variables}
\label{subsec_dimensionless-units}
The most common description of an ultracold Bose gas is that of a wavefunction $\Psi$ which obeys the mean--field Gross--Pitaevskii equation (GPE). In a rotating system such as the one described in Sec. \ref{subsec_EOM}, this equation takes the form 
\begin{equation}
    i \hbar \frac{\partial \Psi}{\partial t} = H_{\Omega} \Psi + g |\Psi|^2 \Psi - \mu \Psi,
    \label{eqn_3D-GPE}
\end{equation}
where  $g = 4\pi \hbar^2 a_s/m$ parameterizes the interaction between multiple particles in the system, $a_s$ is the s--wave scattering length of the particles \cite{dalfovo1999theory}, and $\mu$ is the 3D chemical potential. We are interested in the behaviour of vortices in the rotating plane and so we adopt a highly oblate condensate with trapping frequencies $\omega_{\perp} \ll \omega_\parallel.$ With this tight confinement in the $z$ direction, and the condition $\hbar \omega_\parallel \gg \mu,$ the excitation of modes in the $z$ direction is prevented. This leads to a 3D wavefunction
\begin{equation}
    \Psi_{3D} \left( x,y,z,t\right) = \Psi \left(x,y,t\right) A \exp \left[ - \frac{z^2}{2l_z^2} \right],
    \label{eqn_3d-wavefunction}
\end{equation}
where the $z$ dependence is a Gaussian ground state, and $l_z$ is the oscillator length in the $z$ direction. It is possible to recover a quasi--2D regime by substituting Eqn.~\eqref{eqn_3d-wavefunction} into Eqn.~\eqref{eqn_3D-GPE} and integrating over $z$. In such a quasi--2D system, the interparticle attraction parameter is given by 
\begin{equation}
g_{2D} = \frac{\sqrt{8 \pi} \hbar^2 a_s}{m l_z},
\end{equation} 
and the 2D chemical potential is
\begin{equation}
    \mu_{2D} = \mu - \frac{1}{2} \hbar \omega_\parallel. 
\end{equation}
The GPE for our rotating quasi--2D system is therefore
\begin{equation}
i \hbar \frac{\partial \Psi}{\partial t} = \left( - \frac{\hbar^2}{2m} \nabla^2 + \frac{i\hbar}{m} \boldsymbol{A \cdot } \nabla  + 2m \Omega^2 x^2 + g_{2D} |\Psi|^2 - \mu_{2D} \right) \Psi .
\label{eqn_dimension-gpe}
\end{equation}
This equation is fundamentally different to those of Refs. \cite{Tsubota2002, Kasamatsu2003} as we are in the Landau gauge, given by Eqn.~\eqref{eqn_Landaugauge}. One can convert from the Landau gauge to the symmetric gauge \cite{wood2019quasiperiodic,mingarelli2016simulating} by substituting $\lambda = - m \Omega xy$ into Eqn.~\eqref{eqn:gauge_symmetry}.

We adopt natural units for the system, based on the healing length $\xi = \hbar / \sqrt{m \mu_{2D}}$. This leads to dimensionless distances $x' = x/\xi$ and $y' = y / \xi,$ a dimensionless time $t'= \mu_{2D} t / \hbar,$ and a dimensionless wavefunction $\Psi' = \Psi \sqrt{g_{2D} / \mu_{2D}}.$ 
Using these units we write Eqn.~\eqref{eqn_dimension-gpe} in dimensionless form (dropping the prime notation) 
\begin{equation}
i \frac{\partial \Psi}{\partial t} = H_\Omega \Psi + | \Psi |^2 \Psi - \Psi,
\label{eqn_gpe}
\end{equation}
where the one-body Hamiltonian can be written as
\begin{equation}
H_\Omega = - \frac{1}{2} \nabla^{ 2} + i \Gamma^2 x \frac{\partial}{\partial y} + \frac{1}{2} \Gamma^4 x^{2} ,
\label{eqn_onebody}
\end{equation}
with $\Gamma = \xi / \ell$  the ratio of the healing length $\xi$ to the ``magnetic length'' $\ell$ defined by \cite{yoshioka1983ground, cooper2008rapidly} 
\begin{equation}
\ell^2 = \frac{\hbar}{2m\Omega} .
\label{eqn_ldef}
\end{equation}
In the case of the rotating Bose gas, $\ell$ is a characteristic distance between vortices.


\subsection{Quasi-Periodic Boundary Conditions}
\label{subsec_BCs}

We now wish to consider a representative cell of an infinite rotating system, by introducing quasi-periodic boundary conditions, and to establish the corresponding single-particle basis functions.

For a cell of physical dimensions $0\leq x \leq a \xi$, $0\leq y < b\xi$, with aspect ratio $\kappa  = a/b$, we define our boundary conditions to be (working in dimensionless variables)
\begin{eqnarray}
\Arg \left[ \Psi \left( x+a , y \right) \right] &=& \Arg \left[ \Psi \left( x, y \right) \right] + \frac{2 \pi y}{b} , \label{eqn_xbc} \\
\Arg \left[ \Psi \left( x, y+b \right) \right] &=& \Arg \left[ \Psi \left( x, y \right) \right] .
\label{eqn_ybc}
\end{eqnarray}
Unlike standard periodic boundary conditions, these boundary conditions provide the wavefunction with a winding in the phase which was discovered to be necessary in the work of \cite{byers1961theoretical}. Throughout this paper, we will refer to these boundary conditions as quasi-periodic, or `twisted' \cite{mingarelli2016simulating} boundary conditions. 

From the superfluid velocity in the cell of area $ab$ it is possible to derive a quantisation condition 
\begin{equation}
ab \Gamma^2 = 2 \pi N,
\label{eqn_dimension-quantisation-condition}
\end{equation} 
which relates the area of the cell to the net number of vortices $N$ \cite{cooper2001quantum,Fetter2009rotating}.  With our boundary conditions, the net number of vortices $N$ and the size of the box $a$, $b$ are fixed, and together determine the rotation frequency $\Omega$. Taken together, \eqnreft{eqn_ldef} and \eqnreft{eqn_dimension-quantisation-condition} imply the ``Feynman rule'' of uniform areal vortex density, $n_\mathrm{v}$, mimicking solid-body rotation \cite{Fetter2009rotating}
\begin{equation}
    n_\mathrm{v} = \frac{N}{ab\xi^2} = \frac{m \Omega}{\pi \hbar}.
\end{equation}

We now consider the appropriate basis functions needed to implement a projected Gross-Pitaevskii equation. Previous work \cite{yoshioka1983ground,Schweikhard04,cooper2004vortex, watanabe2004landau, aftalion2005vortex, Sonin05, aftalion2006vortex, Fetter_LLL_2007, Matveenko2009} has investigated rapidly rotating 2D systems which depend only on the Lowest Landau Level (LLL). This is accurate for a system of dense vortices, however where the typical vortex spacing is much larger than the healing length, interactions in the Bose gas lead to contributions from higher Landau levels \cite{cooper2008rapidly}.    Ref.~\cite{yoshioka1983ground} gives the LLL eigenfunction of the Hamiltonian in Eqn.~\eqref{eqn_onebody},  which can be extended to describe higher Landau levels. These eigenfunctions take the form 
\begin{eqnarray}
\phi_{n,k} &=& \sqrt{a \Gamma } \sum_{p=-\infty}^\infty \chi_n \left[ \Gamma a \left( \frac{k}{N} + p \right) - \Gamma x \right] \exp \left[ i \Gamma^2 a \left( \frac{k}{N} + p  \right) y \right], \nonumber \\
\label{eqn_eigenfunction}
\end{eqnarray}
where 
\begin{equation}
\chi_n (x) = \frac{1}{\sqrt{2^n n! \sqrt{\pi}}} H_n (x) \exp \left( - \frac{1}{2} x^2 \right) .
\label{eqn_hermite_function}
\end{equation}
Here, $H_n ( \cdot )$ is the $n^{th}$ physicists' Hermite polynomial \cite{abramowitz1948handbook}, and the Landau levels are indexed by $n=0,1,\dots$. Without loss of generality, we choose to normalise the basis functions to $ab$ (see Appendix~\ref{app_normalisation} for details).
The eigenenergies corresponding to the eigenfunctions of Eqn.~\eqref{eqn_eigenfunction} are 
\begin{equation}
E_{n,k} = \Gamma^2 \left( n + \frac{1}{2} \right) .
\label{eqn_eigenvalue}
\end{equation}

Expanding the wavefunction $\Psi$ in terms of all eigenstates below an energy cutoff $E_\mathrm{cut} = \Gamma^2 (M + 1/2)$ and solving Eqn.~\eqref{eqn_gpe} for the expansion coefficients constitutes the PGPE for this system. The choice of cutoff $M$ will be discussed further in Secs.~\ref{sec_num-method}~and~\ref{sec_convergence-and-testing}.


\section{Numerical Method for Basis Transformation}
\label{sec_num-method}

\subsection{PGPE Implementation}
\label{subsec_rotating-pgpe}
To implement the PGPE for the quasi-periodic system introduced in Sec.~\ref{sec_PGPE}, we follow the same approach as used for the uniform system in Ref.~\cite{Blakie08b}, but using the quasi-periodic one-body eigenstates. As described by Ref.~\cite{Bradley08a}, defining an orthonormal projector with respect to the one-body Hamiltonian is convenient due to the fact the many-body spectrum is well approximated by the single-body spectrum when in the high energy limit. However, in our case there is no known exact numerical quadrature rule for the basis functions with which to implement the projection to numerical precision. Instead we introduce an approximate projection operation that can be made sufficiently accurate for our purposes.

Our basis functions are given by Eqn.~\eqref{eqn_eigenfunction}, and we define the wavefunction $\Psi$ to be
\begin{equation}
\Psi(x,y,t) = \sum_{n=0}^{M-1} \sum_{k=0}^{N-1} c_{n,k}(t) \, \phi_{n,k}(x,y),
\label{eqn_psi-rotating}
\end{equation}
where our energy cutoff is prescribed by the value of $M$, and the summation over $p$ is truncated so that $-p_{max} \leq p \leq p_{max}.$ It is critical that we choose a large enough $p_{max}$ that the quasi--periodic basis functions are approximately orthogonal, and we discuss the validity of this truncation in Sec.~\ref{subsec_truncation-error}. We use the orthonormality conditions of the basis functions (see Appendix~\ref{app_normalisation} for details), to derive an evolution equation for the coefficients $c_{n,k}$
\begin{eqnarray}
i \frac{dc_{n,k}}{dt} &=& \left(E_{n,k} - 1 \right) c_{n,k} \nonumber \\
&+& \sum_{n',m,,m'=0}^{M-1} \sum_{k',j,j'=0}^{N-1} c_{n',k'} c_{m,j} c_{m',j'} \mathcal{I}_{n,n',m,m';k,k',j,j'}
\end{eqnarray}
where 
\begin{equation}
\mathcal{I}_{n,n',m,m';k,k',j,j'} = \int_0^a \int_0^b \phi^*_{n,k} \phi^*_{n',k'} \phi_{m,j} \phi_{m',j'} \, dy dx .
\label{eqn_integral}
\end{equation}
There is no known quadrature rule for the integral in Eqn.~\eqref{eqn_integral}, and so we instead will use an approximate pseudospectral method \cite{boyd}. We write Eqn.~\eqref{eqn_psi-rotating} as 
\begin{equation}
\boldsymbol{\Psi} = T \boldsymbol{c},
\label{eqn_t-transform}
\end{equation}
where $\boldsymbol{\Psi}$ is a real space representation of the wavefunction with $Q^2$ elements indexed by $\boldsymbol{r}_i = (x,y)_i,$ and $\boldsymbol{c}$ is a representation of the wave function in the `coefficient space' of the basis functions, with $MN$ elements indexed by $\boldsymbol{\sigma}_j = (n,k)_j.$ The matrix $T$ is written in terms of the basis functions as 
\begin{equation}
T_{ij} = \phi_{\boldsymbol{\sigma}_j} \left( \boldsymbol{r}_i \right) . 
\label{eqn_t-matrix}
\end{equation}
We must also define the matrix $U$, which is the inverse transformation of Eqn.~\eqref{eqn_t-transform}, i.e. $U = T^\dagger / Q^2,$  and the diagonal `energy matrix' $E$, which contains the eigenvalues of the basis functions, $E_{jj} = E_{\boldsymbol{\sigma_j}}$. The resultant equation for the evolution of the coefficients is 
\begin{equation}
i \frac{d \boldsymbol{c}}{dt} = \left(E - I_{MN} \right) \boldsymbol{c} + U |T \boldsymbol{c} |^2 \left( T \boldsymbol{c} \right) ,
\label{eqn_master-eqn}
\end{equation}
the evolution of which will be discussed in Sec.~\ref{sec_convergence-and-testing}. 

We now consider two sources of error which are unavoidable when performing numerical simulations: the projection error, which arises on choosing the number of grid-points $Q$ for a given $M$, and the error associated with truncating the summation over $p$, which comes from our choice in $p_{max}.$


\subsection{Projection Error}
\label{subsec_projection-error}
As discussed in Sec.~\ref{subsec_rotating-pgpe}, the energy cutoff in our simulations is defined as $M,$ which is the number of Landau levels which are included in our basis functions. We are also working with a system which does not have a quadrature rule, hence there is no clear cut way of selecting a value of $Q$ for a given $M$. The cubic term in the GPE may lead to aliasing in any grid representation of the wavefunction \cite{c-field_review}. In our system, this corresponds to the non-linear term of the GPE producing polynomials of order $3M,$ which are outside the c--field region and hence not energy conserving.  It is therefore necessary to check the validity of any given values of $Q$ and $M$, which we do with the following algorithm.

Assume that our system has $N$ states (vortices), $Q$ grid points in each of the $x$ and $y$ directions, and $M$ Landau levels; for these parameters there is a transformation matrix $T$, and its inverse $U$, the construction of which is described in Eqn.~\eqref{eqn_t-transform}. We generate the matrix $\tilde{T}$ which also has $N$ states and $Q$ grid-points, but has $3M$ Landau levels (on account of the nonlinear term in Eqn.~\eqref{eqn_gpe} being cubic). For the remainder of this section, we use a tilde to denote a coefficient space which has $3M$ Landau levels. 
	
We create a test vector $\tilde{\boldsymbol{c}}$ which is 
	\begin{equation}
	\tilde{\boldsymbol{c}} = \frac{1}{\sqrt{2MN}} ( \overbrace{0, \dots, 0 }^{M \times N} , \overbrace{1, \dots, \dots, 1}^{2M \times N} ). 
	\end{equation}
I.e. the first $M \times N$ elements (which are the coefficients for the basis functions with the lowest $M$ Landau levels) are zero, while the other elements are identical, and normalised so that $|\tilde{\boldsymbol{c}}|^2 =1$. From here, we compute 
\begin{equation}
    \boldsymbol{c} = U \ \left[ \tilde{T} \tilde{\boldsymbol{c}} \right]. 
\end{equation}
This transforms the test vector $\tilde{\boldsymbol{c}}$ from the enlarged basis in coefficient space, into the $Q\times Q$ basis in real space, and then back to the smaller, $M\times N$, coefficient space. 

 Using $\boldsymbol{c}$, the $M\times N$ array of coefficients, we can now quantify the error in the projection. If the projection was perfect, the array $\boldsymbol{c}$ would be precisely zero. That is to say: we would have recovered the coefficients of the lowest $M$ Landau levels from the test array $\tilde{\boldsymbol{c}}$ without alias. 
    
If, however, there are non-zero elements in $\boldsymbol{c}$, then there has been some ``leakage'' of higher order modes into the $M$ lowest modes which we have defined as our c-field. Numerically we define this error to be 
    \begin{equation}
    \delta = \max \{ c_{n,k}^* c_{n,k} \},
    \end{equation}
where this ``leakage'' corresponds directly to momentum aliasing.

\begin{figure}
	\centering
	\includegraphics{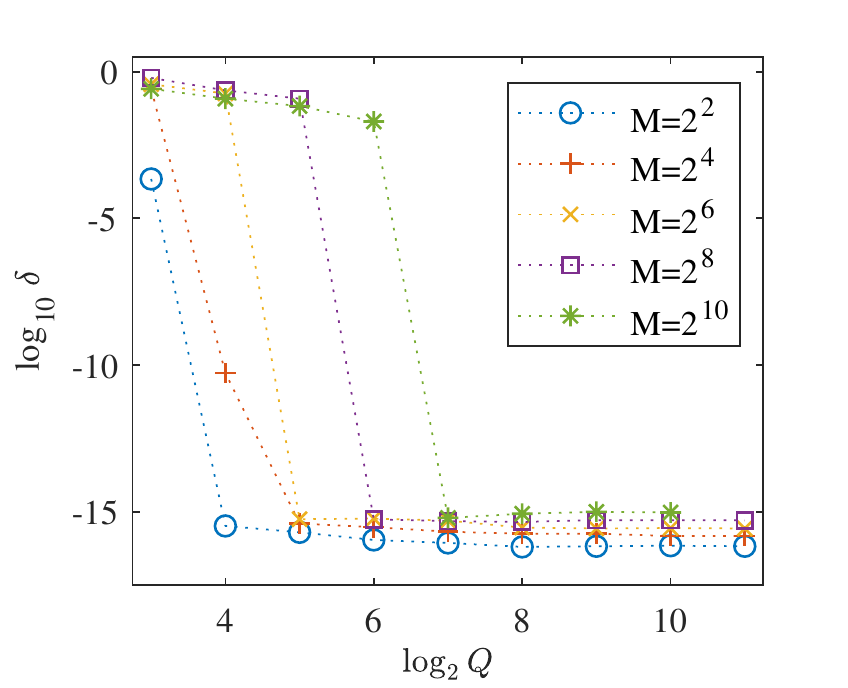}
	\caption{The projection error, $\delta$ as a function of $Q$ for varying values of $M$. We have set $a=b=2^6$ to be the cell size, fixed $p_{max}=10$, and set $N=4$. The dotted lines are added as a guide to the eye.}
	\label{fig:updatedprojectionerror}
\end{figure}

The results of this analysis are presented in Fig.~\ref{fig:updatedprojectionerror}. We see that, for any given $M,$ there is a threshold value of $Q$ for which the projection error $\delta$ becomes negligible. Below these threshold values, the error decreases at a rate which depends on $M$: for small $M$, the error decreases very quickly, while larger $M$ requires more grid-points. Above the threshold value, the projection error converges to a characteristic error for the given set of simulation parameters. This means that increasing the number of points serves only to slow the simulation, and offers no numerical advantage.

We note that the analysis above  was conducted with a cell where $a=b=64$, the truncation $p_{max} =10$, and $N=4$ vortices. A similar analysis can be conducted for a different size cell, and for a different number of states in the system, however we note that the results are qualitatively the same: for higher $M$ one must increase the number of grid-points in order to reduce the projection error. 


\subsection{Truncation Error}
\label{subsec_truncation-error}
Clearly, when calculating the matrix $T$ from the basis functions defined in Eqn.~\eqref{eqn_eigenfunction} it is necessary to truncate the summation over $p.$ We must, however, ensure that we have chosen a large enough value of $p_{max}$ that significant contributions to the wavefunction from neighbouring cells are not erroneously ignored. It is also critical to choose a large enough value of $p_{max}$, as the infinite sum over $p$ is responsible for transforming an integration over a finite domain, into an integration over an infinite domain, which is how the orthonormality of the Hermite polynomials is defined (see Appendix~\ref{app_normalisation} for further details).

There are several well known bounds for the zeros of Hermite polynomials, however the eigenfunctions in Eqn. \eqref{eqn_eigenfunction} are a sum over a product of a Hermite function $\chi_n(x)$, and the complex exponential in $y$. Although Hermite functions decay exponentially quickly after their most extreme zeros, there is still an imaginary part of these eigenfunctions which must be taken into account. The presence of $p$ in both the $x$ and $y$ components of the basis functions mean that truncating the summation over $p$ is not as simple as using a bound for the Hermite polynomials, and we must be cautious that the value of $p_{max}$ is chosen correctly.  

\begin{figure}
	\centering
	\includegraphics[width=\linewidth]{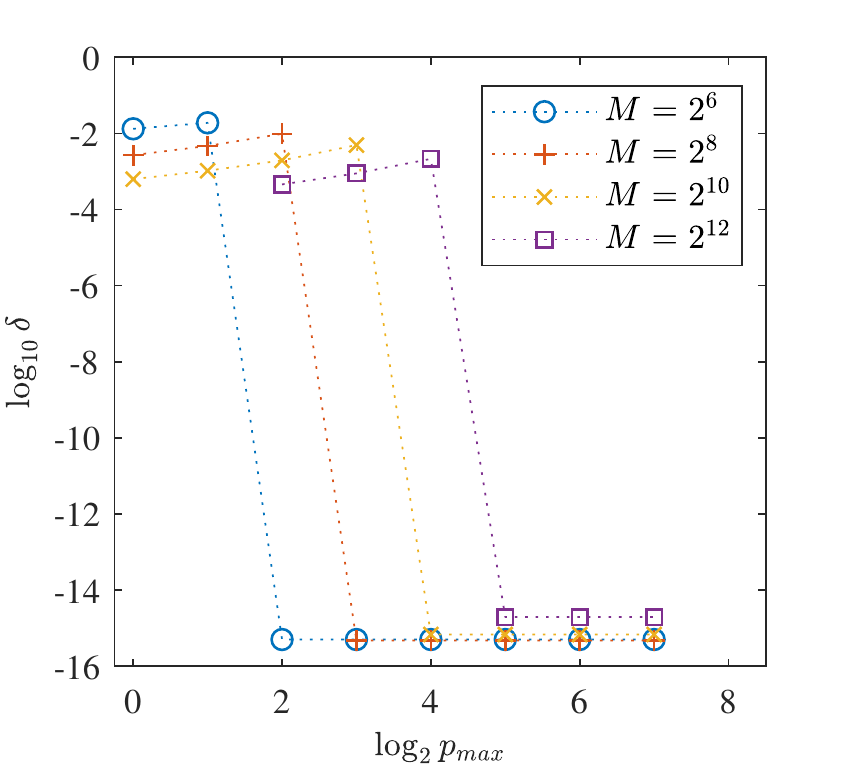}
	\caption{The truncation error $\delta$, for varying $p_{max}$ with fixed $M$ and $Q$. For $M \in \{ 2^6, 2^8, 2^{10} \}$ we used $Q = 2^7$ grid-points, while for $M>2^{10}$, it is necessary to use $Q=2^8$  grid-points to achieve a meaningful result. Note that $a=b=2^6$ and $N=4$ in this analysis. The dotted lines are added as a guide to the eye.}
	\label{fig:pmaxvary}
\end{figure}

We perform the same analysis as in Sec.~\ref{subsec_projection-error} in order to quantify the error $\delta,$ however in each case we fix $Q$ and $M$ and instead vary $p_{max}$. The results can be found in Fig.~\ref{fig:pmaxvary}. For each $Q$ and $M$, we note there is a threshold value of $p_{max}$ above which the truncation error becomes negligible (this is indicated by a sudden drop in the value of $\delta$ in Fig.~\ref{fig:pmaxvary}). Initially there is an increase in the error (for $p_{max}=1$), however this is because the basis functions do not converge to the correct value for this choice in truncation. Above the threshold value, there is a convergence in the error for a given $M$ and $Q$. 

Informed by the analysis of Sec.~\ref{subsec_projection-error}, we note that for values of $M$ which were greater than $2^{10}$, it was necessary to use $Q=2^8$, grid points in each direction to get a meaningful result.

\section{Vortex Ansatz for Initial Condition}
\label{sec_vortex-ansatz}

In this section we describe the process by which we prepare an initial configuration of $N_\mathrm{v}$ vortices placed within the cell. This allows us to investigate a number of scenarios involving free vortices, clustered vortices and dipole pairs. 
\begin{figure*}
	\centering
	\includegraphics[width=\linewidth]{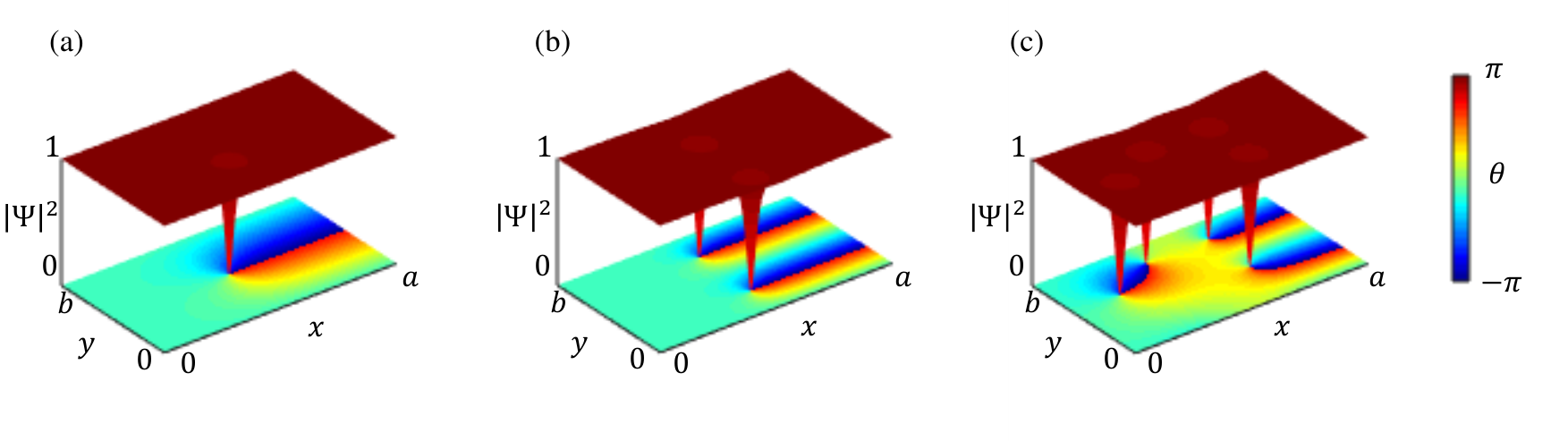}
	\caption{Example configuration of vortices using the method described in Sec. \ref{sec_vortex-ansatz}. The colour bar indicates the phase of the superfluid. (a) A single, positively charged, vortex is placed at the centre of the cell. (b) Two positively charged vortices are placed at $(a/2,3b/4)$ and $(a/2,b/4)$. (c) Three positively charged and one negatively charged vortices create a dipole pair in the cell.}
	\label{fig:vortexphase}
\end{figure*}

It is known that it is possible to express the phase of a vortex using the zeros of a Weirstrass function \cite{Tkachenko1966lattice}. Further, in the Landau gauge it is appropriate to use Jacobi Theta functions to describe the phase. The 3rd Jacobi Theta Function is defined as \cite{abramowitz1948handbook}
\begin{equation}
\vartheta_3 \left( z , \tau \right) = 1+ 2 \sum_{n=1}^\infty  q^{ n^2} \cos \left( 2 n z \right), 
\label{eqn_theta3}
\end{equation}
where $z$ is a complex coordinate, and $\tau \in \mathbb{C} $ is the lattice parameter with nome $q=\exp \left( i \pi \tau \right).$ We restrict ourselves to the case of a rectangular domain, requiring $\Re \left( \tau \right) = 0$ and $\Im \left( \tau \right) > 0,$ so that $\vartheta_3$ has quasi-periodicity relation  
\begin{equation}
\Arg \left[ \vartheta_3 \left( z + f\pi + g \tau \pi ; \tau \right) \right] = \Arg \left[ \vartheta_3 \left( z; \tau \right) \right]  - 2g \Re (z)  , 
\label{eqn_theta-3-qp}
\end{equation}
for integers $f$ and $g$. In order to describe a domain which is arbitrary sized, we introduce $L$ then by re-scaling $z \to \pi z / L,$ and defining the lattice parameter $\tau $ to be purely imaginary, the Jacobi theta function $\vartheta_3$ is quasi periodic on $0\leq \Re\left(z\right)<L$ and  $0\leq\Im\left(z\right)<L\Im(\tau).$ In this case,  the quasi-periodicity relation of Eqn.~\eqref{eqn_theta-3-qp} becomes
\begin{equation}
\Arg \left[ \vartheta_3 \left( \frac{\pi}{L} \left( z+L \tau \right) ; \tau  \right) \right] = \Arg \left[ \vartheta_3 \left( \frac{\pi z}{L} ; \tau \right)  \right] - \frac{2\pi}{L} \Re (z).
\label{eqn_eqn43}
\end{equation}
 By comparison with the quasi-boundary conditions of Eqn.~\eqref{eqn_xbc}, it follows that $L = b$, $\tau=i\kappa$ and $z=ix-y$. 
 Consequently, it is possible to determine that the fundamental solution for the phase $\theta$ of a vortex centred in the box at $(a/2, b/2),$ is 
 \begin{equation}
     \theta(z) = c \Arg \left[ \vartheta_3 \left( \frac{\pi}{b} z \, ; \, i \kappa \right) \right],
 \end{equation}
 where $c$ is the integer charge of the vortex.
 This fundamental solution is shown in the phase plot of Fig.~\ref{fig:vortexphase} (i). By the use of a suitable gauge transformation, it can be shown that this is equivalent to expressions obtained for quasi-periodic boundary conditions in the symmetric gauge in Ref.~\cite{wood2019quasiperiodic}.
 
 Suppose that we wish to obtain the phase of the $k$th vortex, of charge $c_k$, which is shifted from the centre of the cell, to the position $\left( x_{k}, y_{k} \right)$. Then we define the effective vortex coordinate 
 \begin{equation}
     z_{k} = i \left( x_{k} - \frac{a}{2} \right) -  \left( y_{k} - \frac{b}{2} \right),
 \end{equation} so that the phase of the $k$th vortex is given by
 \begin{equation}
     \theta_k \left( z \, ; \, z_{k} \right) = c_k \Arg \left[ \vartheta_3 \left( \frac{\pi}{b} [z - z_{k} ] ; i \kappa \right) \right].
 \end{equation}
 
 The density profile of a vortex was found numerically in Ref.~\cite{Bradley2012a}. Non-dimensionalising this function, and setting the background density to be one, we have
 \begin{equation}
     \rho_k \left( z \, ; \, z_{k}\right) = \left[ \frac{\big|z-z_{k} + \frac{1}{2}(ia-b) \big|^2}{\big|z-z_{k} + \frac{1}{2}(ia-b) \big|^2 + \Lambda^{-2}} \right]^{1/2},
 \end{equation}
 where $\Lambda \approx 0.8249$ is a universal constant.
 
 Combining phase and density profiles of the individual vortices, our ansatz wavefunction $N_\mathrm{v}$ vortices is
  \begin{equation}
      \Psi \left( z \, | \, \{z_k\} \right) = \prod_{k=0}^{N_\mathrm{v}-1} \rho_k \left( z ; z_{k} \right) \exp \left[ i \theta_k \left( z; z_{k} \right) \right],
      \label{eqn_wavefunction-ansatz}
  \end{equation}
  where $ \{z_k\} = \{ z_0, \ldots, z_{N_\mathrm{v}-1} \}$.
  In order to determine the symmetry conditions of this ansatz, let us consider the transformation $x\to x+a$. In this case, we have 
  \begin{equation}
   \Arg \left[ \Psi \left( z + ia \, | \, \{z_k\} \right) \right] = \sum_{k=0}^{N_\mathrm{v}-1} c_k \Arg \left[ \vartheta_3 \left( \frac{\pi}{b} \{ z -z_{k} \} + i \frac{\pi a}{b} ; i \kappa \right) \right],   
  \end{equation}
  which, using the quasi-periodicity relation of Eqn.~\eqref{eqn_theta-3-qp}, is 
  \begin{eqnarray}
   \Arg \left[ \Psi \left( z + ia \, | \, \{z_k\} \right) \right] &=& \nonumber \\ 
   \Arg \left[ \Psi \left( z \, | \, \{z_k\} \right) \right] &+& \frac{2 \pi N y}{b} + \frac{\pi}{b} \sum_{k=0}^{N_\mathrm{v}-1} \left( y_{k} - \frac{b}{2} \right),   
   \label{eqn_shifted_phase}
  \end{eqnarray}
  where $N$ is the net number of vortices (the sum of $c_k$). The first two terms on the right hand side of Eqn.~\eqref{eqn_shifted_phase} are in direct agreement with the quasi-periodic boundary conditions of Eqns.~\eqref{eqn_xbc}~and~\eqref{eqn_ybc}. However, to match the boundary conditions the third term must vanish. This means that the vortex positions $y_k$ must satisfy
  \begin{equation}
  \bar{y_v} = \frac{1}{N} \sum_{k=0}^{N_\mathrm{v} -1} c_k y_{k} = \frac{b}{2},
  \end{equation}
  placing the center of vorticity at $b/2$ in the $y$-direction.
  This condition is related to the fact that the ground state vortex lattice breaks the translational symmetry of the system. Adding a constant to our boundary conditions [Eqn.~\eqref{eqn_ybc}] would trivially shift the center of vorticity within the cell. An equivalent connection between boundary conditions and the center of vorticity is found for quasi-periodic boundary conditions in the symmetric gauge \cite{wood2019quasiperiodic}. 
  Fig.~\ref{fig:vortexphase} shows a small selection of initial vortex configurations which can be created using the ansatz wavefunction of Eqn.~\eqref{eqn_wavefunction-ansatz}.

\section{Convergence and Testing of the Method}
\label{sec_convergence-and-testing}
\subsection{Overview of Numerical Procedure}
Here we briefly outline how the pseudospectral method described above can be implemented numerically. 
In order to perform the transformations between real and coefficient space required by Eqn.~\eqref{eqn_t-transform}, we begin by creating the matrix described in Eqn.~\eqref{eqn_t-matrix}. Note that this fixes the dimensions of the fundamental cell, $a,b$ and $\kappa$, the number of Landau levels, $M$, the number of grid-points, $Q$, and the net number of vortices, $N$. Once this is complete, we evolve Eqn.~\eqref{eqn_master-eqn} from an initial condition. Numerically, we compute the time evolution using an adaptive 8th order Dormand Prince (DP8) method \cite{prince} with adaptive time stepping subject to an error tolerance $\epsilon$. Since the majority of the memory requirements lie in the storing of the $T$ and $U$ matrices, the extra memory required to use such a high order time-stepping scheme is inconsequential. The high order of the method reduces the total number of time derivative evaluations required while maintaining sufficiently stringent tolerance to preserve the conserved quantities to good accuracy over long time. The most computationally demanding step in the procedure is performing the basis transformations needed to evaluate the time derivative; this amounts to performing multiplication by the matrices $T$ and $U$, which have a large size of $MNQ^2$ elements (about $2^{27}$ for typical parameters). Owing to the large size and high condition number of the $T$ and $U$ matrices, numerical rounding errors in these matrix-vector multiplications can become non-negligible with standard double-precision arithmetic. We find that performing a stabilized matrix-vector multiplication, using the techniques to extend precision described in Ref.~\cite{Dekker} and parallelized using OpenMP, effectively eliminates these problems without significantly increasing computation times.

There are two kinds of initial conditions that we may use. In the first instance, we can control the occupation of the modes in coefficient space, in a manner similar to the simulations of Ref.~\cite{gasenzer2012}. More conveniently, we can produce an ansatz wavefunction whereby we prescribe the position and charge of $N$ vortices, using the method described in Sec.~\ref{sec_vortex-ansatz}. The only difference is that we must transform this ansatz into coefficient space before evolving.

\subsection{Conserved Quantities}
There are three quantities which should be conserved by any numerical treatment of Eqn.~\eqref{eqn_gpe}. They are the real-space norm $\mathcal{N}_R$ of the wavefunction,
	\begin{equation}
	 \mathcal{N}_R (t) = \int_0^a \int_0^b \Psi^* (x,y,t) \Psi(x,y,t) \, dydx,
	 \label{eqn_real-space-norm}
	\end{equation}
	the norm of the coefficients, $\mathcal{N}_C,$ defined as
	\begin{equation}
	\mathcal{N}_C (t) = \sum_{j=0}^{MN-1} c^*_{\boldsymbol{\sigma}_j} (t) \, c_{\boldsymbol{\sigma}_j}(t),
	\end{equation}
    and the energy of the system, 
    \begin{eqnarray}
    \mathcal{E}(t) &=& \frac{1}{\mathcal{N}_C(0)}   \sum_{j=0}^{MN-1} E_{\boldsymbol{\sigma}_j} c^*_{\boldsymbol{\sigma}_j} (t) \, c_{\boldsymbol{\sigma}_j}(t) \nonumber \\
    &+& \frac{1}{\mathcal{N}_R (0)} \int_0^a \int_0^b \frac{1}{2} | \Psi(x,y,t) |^4  \, dydx.
    \label{eqn_energy}
    \end{eqnarray}
    In both Eqns.~\eqref{eqn_real-space-norm}~and~\eqref{eqn_energy} we have discretized real space, and so the integrals will be replaced with summations, with $dx \to a/Q$ (likewise $dy \to b/Q$). Due to numerical error, these quantities will not be conserved by our evolution scheme. Tracking their changes, however, provide a key insight as to how accurate our scheme is.


\subsection{Evolution of Vortex Ansatz States}
\label{subsec_ansatz-states}
\begin{figure*}
	\centering
	\includegraphics[width=\linewidth]{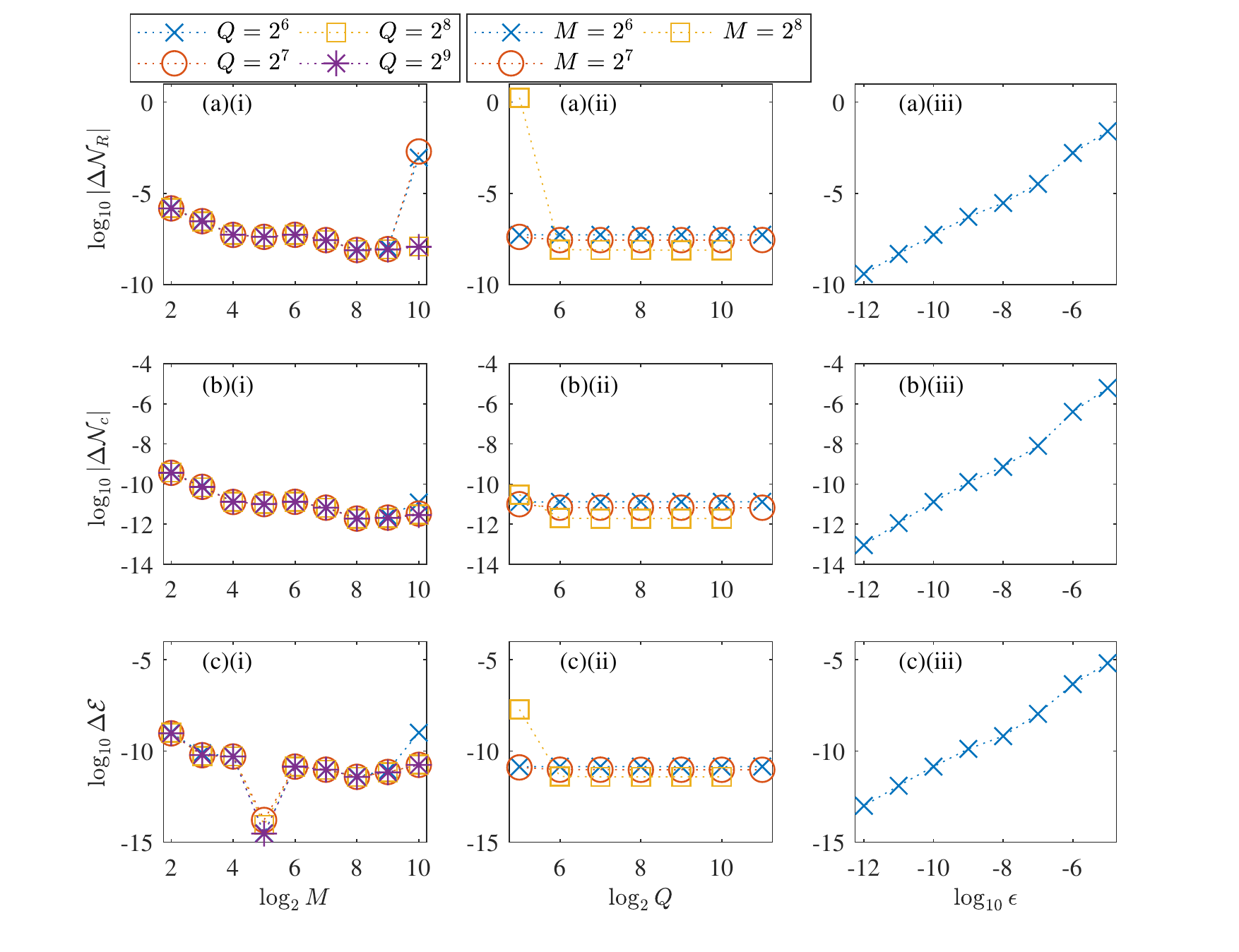}
	\caption{Evolution error for the quantities $\mathcal{N}_R$, row (a); $\mathcal{N}_C$, row (b); and $\mathcal{E}$, row (c). Column (i): varying $M$ for $Q=2^6$, blue crosses; $Q=2^7$, red circles; $Q=2^8$, yellow squares; $Q=2^9$, purple asterisks. Column (ii): varying $Q$ for $M=2^6$, blue crosses; $M=2^7$, red circles; $M=2^8$, yellow squares. For columns (i) and (ii), $\epsilon=10^{-10}$. Column (iii): varying $\epsilon$ for $M=2^6$ and $Q=2^8$. In all cases, $a=b=2^6$. }
	\label{fig:updatedevolutionerror}
\end{figure*}

We begin with a state wich is a random configuration of $N=4$ vortices, in a square cell with side lengths $a=b=64.$ This initial state is then evolved to $t_{final} = 50$ (in dimensionless time units), and the difference between the initial and final values are computed, i.e. $\Delta \mathcal{N}_R = \mathcal{N}_R (0) - \mathcal{N}_R (50).$ The results of this can be seen in Fig.~\ref{fig:updatedevolutionerror}.

In column (a)(i)--(c)(i), we calculate the evolution error for varying values of $M,$ while the tolerance in the numerical timestepping is fixed, $\epsilon=10^{-10}$. We do this for a number of different grid points: $Q=2^6$, blue crosses; $Q=2^7$, red circles; $Q=2^8$, yellow squares; $Q=2^9$, purple asterisks. We note that the curves have a characteristic bow shape; initially increasing the number of Landau levels decreases the error in the evolution. For each value of $Q$, however, there comes a point where projection error dominates the increase in $M$, and the evolution error increases. This is particularly noticeable in the regime of low $Q$ and high $M$ in the plot of $\Delta \mathcal{N}_R$, Fig.~\ref{fig:updatedevolutionerror}~(a)(i). 

In column (a)(ii)--(c)(ii), we calculate the evolution error for varying values of $Q$ for a fixed tolerance of $\epsilon=10^{-10}$, with $M=2^6$, blue crosses; $M=2^7$, red circles; $M=2^8$, yellow squares. We observe that increasing the number of grid points $Q$ leads to a monotonic decrease in the evolution error. Initially projection error dominates, however this is in a regime where we have one or fewer grid points per healing length. As $Q$ increases beyond approximately 4 grid points per healing length, we note that the error converges for each value of $M$; it it also apparent that once the error has converged, a higher value of $M$ leads to a better conservation in the quantities of interest.  

In column (a)(iii)--(c)(iii), we calculate the evolution error for varying values of $\epsilon$, where $M=2^6$ and $Q=2^8$. We see that there is a very good agreement between the tolerance size, and the expected error of the DP8 method.

It should be noted that although this demonstrates the evolution error of one initial state, it is qualitatively representative of all initial states. That is to say, the results of the evolution error testing presented here are a realisation of a single initial condition, but we note that this is indicative of all initial conditions.


\subsection{Stability of the Ground State}
\label{subsec_ground-states}
As well as performing the dynamical evolutions described in the previous sections, we want to be able to find the ground state of a system with $N$ vortices.  In order to do this, we add a dimensionless damping parameter $\gamma$ to the governing equation \cite{Landau&Lifshitz, Tsubota2002}. This parameter describes the diffusion of thermal atoms from the system, a key physical process in relaxing the system to a ground state \cite{billam2014prl}. This means that Eqn.~\eqref{eqn_gpe} becomes 
\begin{equation}
i \frac{\partial \Psi}{\partial t} = \left( 1 - i \gamma \right) \left[ H_\Omega \Psi + | \Psi |^2 \Psi - \Psi \right],
\label{eqn_damped-gpe}
\end{equation}
and hence we will numerically simulate
\begin{equation}
 \frac{d \boldsymbol{c}}{dt} = \left(  \gamma -i \right) \left[ \left(E-I_{MN}\right)\boldsymbol{c} + U | T\boldsymbol{c}  |^2 (T\boldsymbol{c}) \right].
\label{eqn_dpgpe}
\end{equation}
For a domain with aspect ratio $\kappa = \sqrt{3}$, the ground state has been shown to be a hexagonal lattice \cite{komineas2012vortex,Tkachenko1966lattice, abrikosov1957magnetic}. We will show in the rest of this section that this damped PGPE will cause the system to relax into a vortex lattice ground state. 

\begin{figure}
    \centering
    \includegraphics{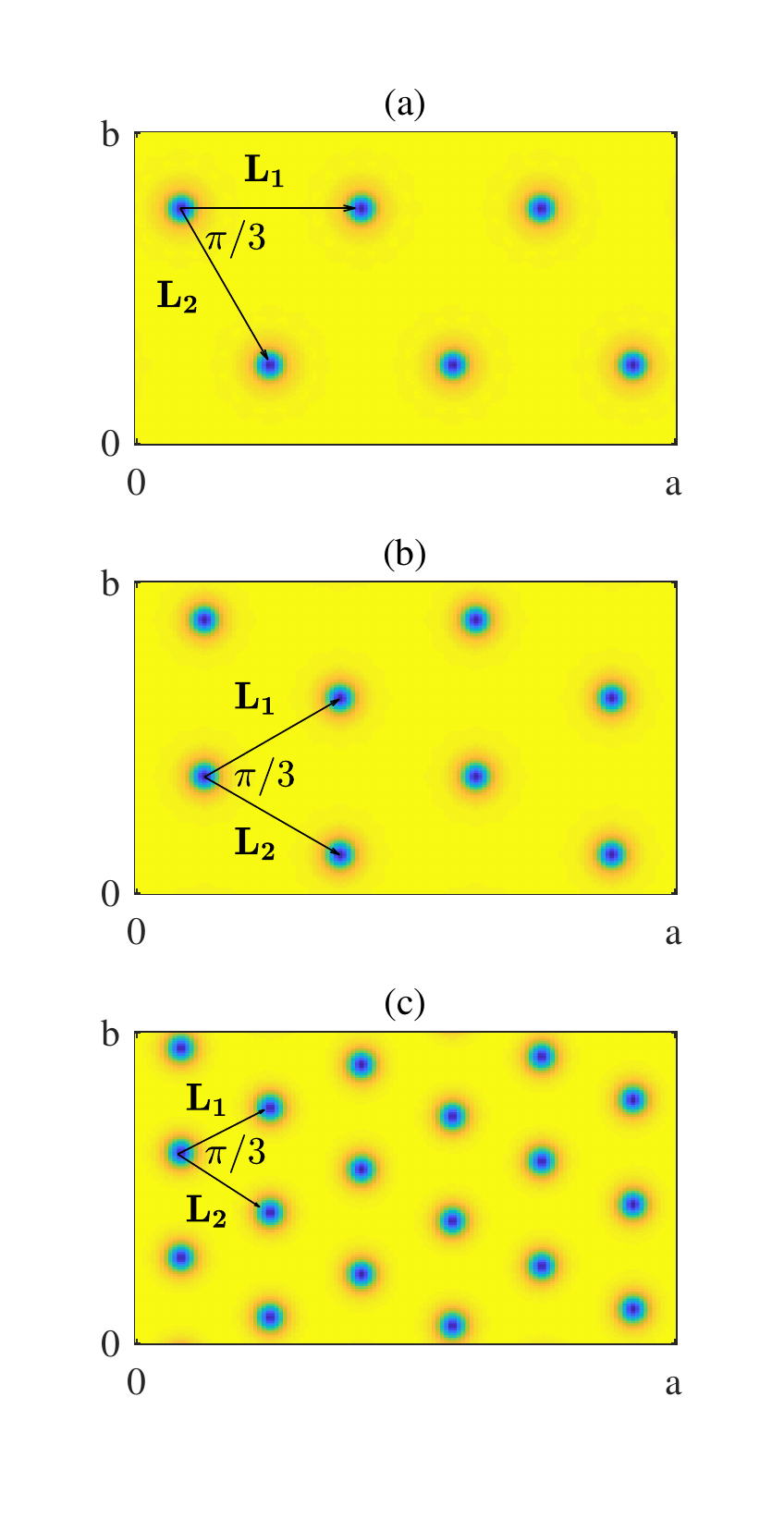}
    \caption{The hexagonal lattice ground states. (a) a system with $N=6$ vortices, (b) a system with $N=8$ vortices, and (c)  a system with $N=18$ vortices. The primitive vectors of a hexagonal lattice, $\boldsymbol{L}_1$ and $\boldsymbol{L}_2$, are added as a guide to the eye. In each case, $a=32\sqrt{3}$ and $b=32$.}
    \label{fig:hexagonal_groundstate}
\end{figure}

The procedure is as follows: We initially seed all of the coefficients so that 
\begin{equation}
    c_{n,k}(0) = \frac{(1+i)}{\sqrt{2NM}},
\end{equation} and evolve this state using the damped GPE in Eqn.~\eqref{eqn_dpgpe}, with the parameter $\gamma=1$. This leads to the ground state $\boldsymbol{c}^{(g)}$. In Fig.~\ref{fig:hexagonal_groundstate} we plot the ground state for $N=6$, $N=8$ and $N=18$.

A lattice is characterised by a pair of primitive lattice vectors $\boldsymbol{L}_1$ and $\boldsymbol{L}_2$, from which we can infer the shape of a lattice (i.e. square, hexagonal, etc.). In Fig.~\ref{fig:hexagonal_groundstate} we add the primitive vectors of a hexagonal lattice, such that $|\boldsymbol{L}_1| = |\boldsymbol{L}_2|$ and $\boldsymbol{\hat{L}}_1 \boldsymbol{\cdot \hat{L}}_2 = 1/2$, confirming that the ground state is a hexagonal lattice. Further, we observe that in the long term the energy of the system is monotonically decreasing when evolving Eqn.~\eqref{eqn_damped-gpe} with $\gamma=1$, and that the energy converges. For the parameters in Fig.~\ref{fig:hexagonal_groundstate}, $\mathcal{E}(t+\delta t) - \mathcal{E}(t)$ has converged to within at least $2\times 10^{-7}$.

\section{Application: Lattice Melting}
\label{sec_application}
Here we present an application of the method to simulate a melting vortex lattice. Evolving an initial configuration of 6 vortices using the damped GPE leads to a lattice ground state $\boldsymbol{c}^{(g)}$,  as reported in Sec.~\ref{subsec_ground-states}.

We then add noise to the ground state, by taking 
\begin{equation}
    c_{n,k} =  \eta c_{n,k}^{(g)} + (1-\eta) \exp \left[ i \varpi \right]
\end{equation}
for $n=1,\ldots,(M-1)$, where the parameter $\eta$ controls the amount of noise which is injected into the lattice ground state, and $\varpi$ is sampled from a uniform distribution $U(0,2\pi)$. Adding noise to the coefficients of the ground state will increase the presence of higher Landau levels in the system, and hence effect the thermal properties of the system.  

\begin{figure*}
\centering
\includegraphics[width=\linewidth]{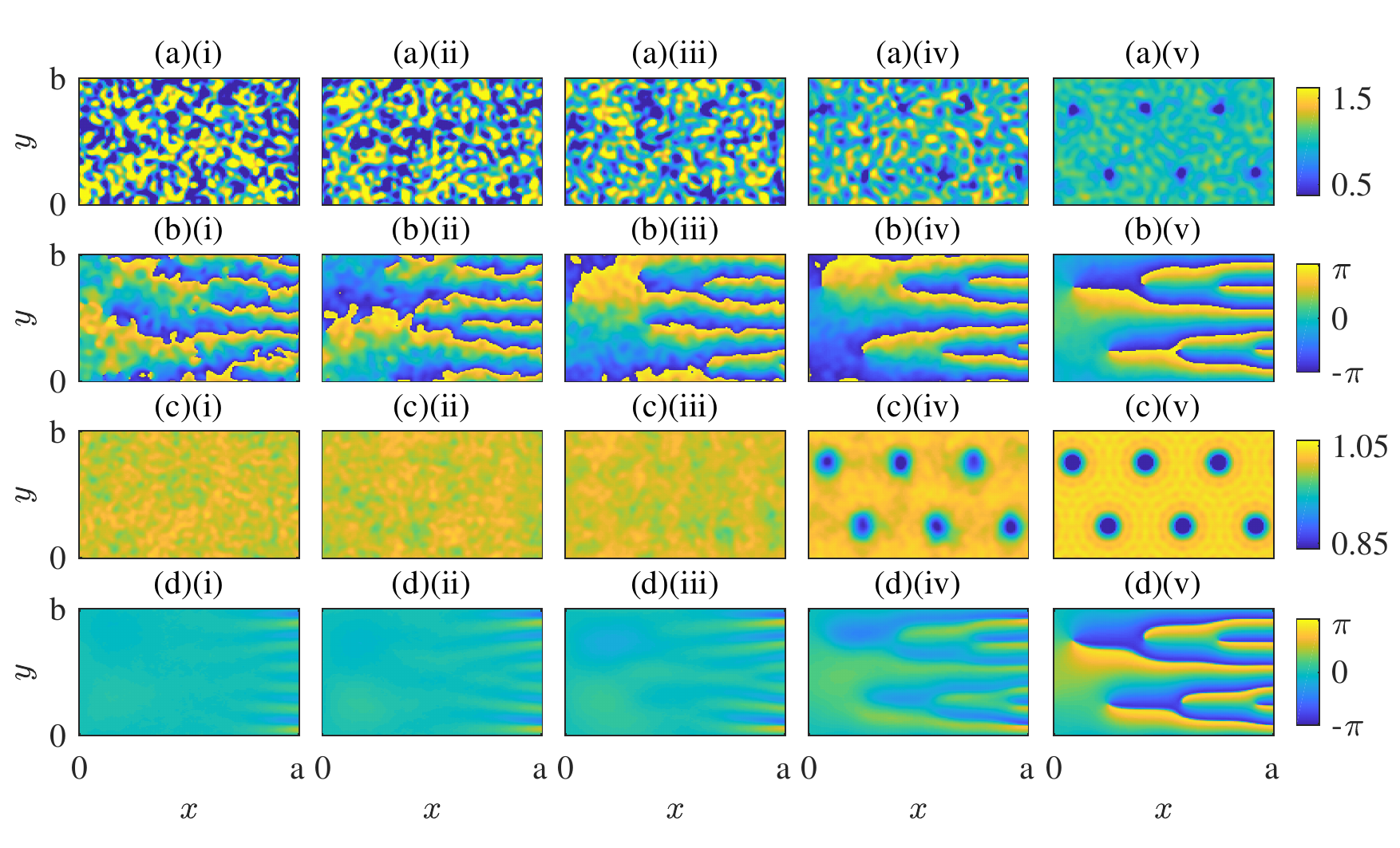}
\caption{Row (a)(i)--(a)(v): instantaneous density profile at $t=5000$. Row (b)(i)--(b)(v): instantaneous phase profile at $t=5000$. Row (c)(i)--(c)(v): time and ensemble averaged density profile, $\bar{\rho}$. Row (d)(i)--(d)(v): time and ensemble averaged phase profile, $\bar{\theta}$.  The initial configurations are given by: Column (a)(i)--(d)(i): $\eta=0.982$, column (a)(ii)--(d)(ii): $\eta=0.986$, column (a)(iii)--(d)(iii): $\eta = 0.990$, column (a)(iv)--(d)(iv): $\eta=0.994$, and column (a)(v)--(d)(v): $\eta = 0.998$. See Supplemental Material \cite{link} which contains movies of the time evolution. }
\label{fig:lattice_melting}
\end{figure*}

Here we  take 5 values of $\eta$, so that the initial configuration is  98.2\%, 98.6\%, 99\%, 99.4\% and 99.8\% of the lattice ground state. For each of these configurations, we simulate 10 different realisations of noise added to the coefficients of the ground state, evolved to dimensionless time $t_{f} = 10^4$. In addition to the individual trajectories, we compute the time and ensemble averaged density,
\begin{equation}
    \bar{\rho} = \frac{1}{t_f - t_i} \int_{t_i}^{t_f} \Big\langle \big| \Psi \left(x,y,t \right) \big|^2 \Big\rangle \ dt,
\end{equation}
and the time and ensemble averaged phase. Computed numerically over $R$ trajectories, this is 
\begin{equation}
    \bar{\theta} = \Arg \left[ \prod_{t=t_i}^{t_f} \prod_{r=1}^{R} \exp \left( i \frac{\Arg \left[\Psi_r(x,y,t)\right] - \Arg \left[\Psi_r(0,0,t) \right]}{R \left(t_f - t_i \right)} \right) \right]. 
\end{equation}
We compute these averages over an ensemble of 10 trajectories, averaging in time from $t_i=5\times 10^3$ to $t_f=10^4$, numerically integrated over $500$ equally-spaced outputs. Although we do not compute the temperatures that these energies correspond to in the microcanonical ensemble, in principle these can be determined as described by Ref.~\cite{Blakie05a}.

Fig.~\ref{fig:lattice_melting} shows the instantaneous and averaged density and phase profiles for the different values of $\eta$.  For reference, the energy of the lattice ground state is  $\mathcal{E}_g = -0.7135$.  Due to the degeneracy of eigenenergies, the parameter $\eta$ is not a versatile measure of the injected energy for systems with different numbers of vortices. Further, the initial energy of each realisation is different, and so we compare different values of noise in the system by computing the added energy, $\mathcal{E}_A = \langle \mathcal{E}_0 \rangle - \mathcal{E}_g$, where $\langle \mathcal{E}_0 \rangle$ is the energy of the system after one time step, so that the wavefunction and vector of coefficients is correctly normalised.  In Fig.~\ref{fig:lattice_melting}, column (i) corresponds to $ \mathcal{E}_A = 0.8107 \, |\mathcal{E}_g| $, column (ii) corresponds to $ \mathcal{E}_A = 0.6688 \, |\mathcal{E}_g| $, column (iii) corresponds to $ \mathcal{E}_A = 0.5361 \,  |\mathcal{E}_g| $, column (iv) corresponds to $ \mathcal{E}_A = 0.4267 \, |\mathcal{E}_g| $, and column (v) corresponds to $ \mathcal{E}_A = 0.3638 \, |\mathcal{E}_g| $.

It is clear to see that as the energy of the system increases, stronger fluctuations destroy the regular vortex lattice. In Fig.~\ref{fig:lattice_melting}~(a)(i)--(b)(i) we see that fluctuations have led to the creation of short--lived dipole pairs, which in turn means that there is no recognisable structure to the time and ensemble avearged profiles, Fig.~\ref{fig:lattice_melting}~(c)(i)--(d)(i). Similarly, fluctuations in Fig.~\ref{fig:lattice_melting}~(a)(ii)--(b)(ii) prevent the formation of a lattice in Fig.~\ref{fig:lattice_melting}~(c)(ii)--(d)(ii)

In Fig.~\ref{fig:lattice_melting}(a)~(iv)--(d)(iv), we see that while the instantaneous density profile, Fig.~\ref{fig:lattice_melting}(a)(iv), contains sharp fluctuations, a hexagonal vortex lattice endures in the averaged density profile, Fig.~\ref{fig:lattice_melting}~(c)(iv). Here the edges of the vortex cores appear fainter than in the lattice of Fig.~\ref{fig:lattice_melting}~(c)(v), due to oscillations in the position of the vortices in individual trajectories. Indeed, the main difference between the averaged density profiles of Figs.~\ref{fig:lattice_melting}~(c)(i) -- (c)(v) is that the lattice melts as the system becomes dominated by fluctuations, which is the component of the thermal cloud that exists within the classical region \cite{Blakie08b}.

In the ensemble with the smallest additional energy, Fig.~\ref{fig:lattice_melting}~(a)(v)--(d)(v),  we see that even in instantaneous profiles, Figs.~\ref{fig:lattice_melting}~(a)(v) and (b)(v), the vortex lattice is preserved. Indeed, the fluctuations due to this small amount of injected energy are highly smoothed out by time and ensemble averaging [Figs.~\ref{fig:lattice_melting}~(c)(v) and (d)(v)] so that we recover profiles similar to the ground state of Fig.~\ref{fig:hexagonal_groundstate}~(a).

\begin{figure}
    \centering
    \includegraphics{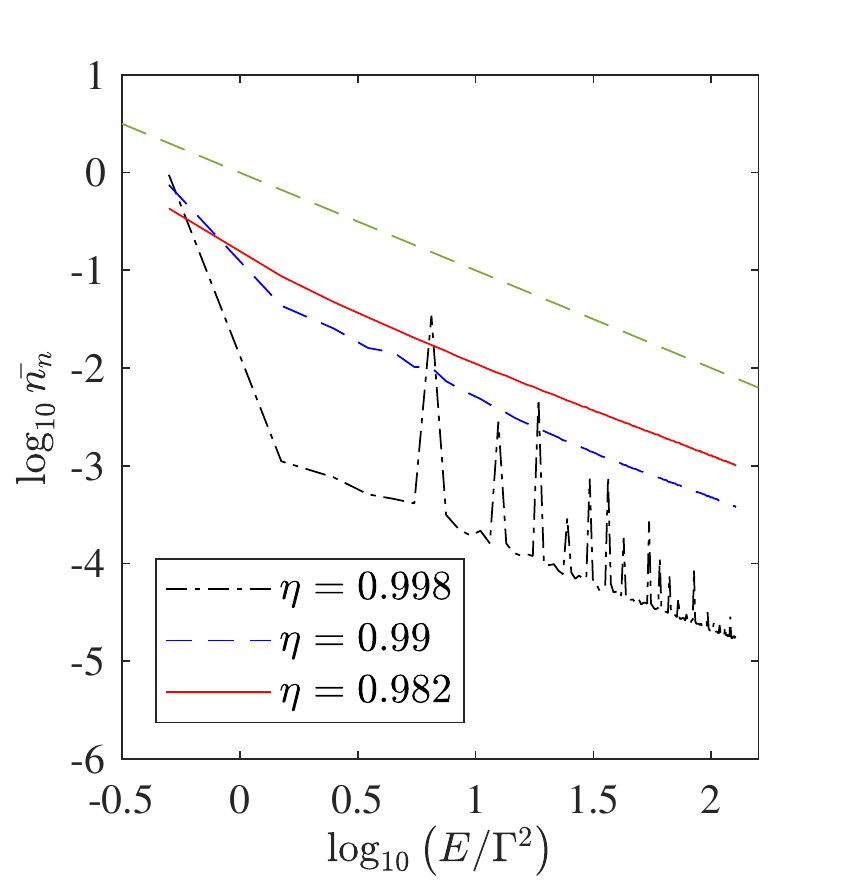}
    \caption{The time and ensemble averaged occupation of $c_{n,k}$ as a function of Landau levels for $\eta = 0.982$, solid red line, $\eta=0.990$, dashed blue line, and $\eta=0.998$, dot-dashed black line. The equipartition of energy, $1/E$, green dashed line, is added as a guide to the eye.}
    \label{fig:landau_level_filling}
\end{figure}

In Fig.~\ref{fig:landau_level_filling} we plot the time and ensemble averaged occupation of the Landau levels. Here we define
\begin{equation}
    \bar{n_n} = \frac{1}{t_f - t_i} \int_{t_i}^{t_f} \sum_{k=0}^{N-1} \ \Big\langle | c_{n,k} ( t ) |^2 \Big\rangle \ dt, 
\end{equation}
as the index of the state (vortex) does not enter into the expression of eigenenergies. We notice that, by adding enough noise to the ground state (corresponding to a low value of $\eta$), the distribution of Landau level occupation is proportional to $1/E$, which corresponds to classical equipartition of energy over the modes. For a high value of $\eta$,  although the majority of the Landau level occupation is centered around the lowest Landau levels, the effects of rotation on the system cause the formation of some structure in the filling of higher modes corresponding to the vortex lattice. The value of $\eta=0.990$ represents a crossover between these limits. A large proportion of the filling is in the Lowest Landau levels, indicating the presence of a condensate. However, higher modes are still significantly occupied, destroying the lattice structure, and indicating the presence of thermal effects.

\section{Conclusion and Outlook}
\label{sec_conclusion}

In this paper we have presented an efficient method for simulating a harmonically trapped Bose gas, which is rotating at the centrifugal limit. We have shown that it is possible to do so without the issue of edge effects by choosing suitable quasi-periodic boundary conditions. We have used the single particle basis functions with these quasi-periodic boundary conditions to expand the wavefunction below an energy cutoff, thus implementing the PGPE for this rotating system. There are two primary sources of error which arise from such a pseudospectral method in this case; these are the error in projection caused by choosing an energy cut-off, $M$, and the error associated with truncating an infinite summation appearing in the basis functions themselves. We have quantified these errors, and have shown that for suitable choices of simulation parameters it is possible to reduce these errors to an acceptably small value.

On adding damping, our PGPE relaxes non-equilibrium initial conditions to the expected regular vortex lattice ground state. It is also extremely useful to be able to set up an initial condition composed of $N_\mathrm{v}$ vortices with arbitrary integer charge placed at any points in the domain (subject to symmetry conditions); we have given an ansatz wavefunction for such an initial condition, working in the Landau gauge. Finally, as an application of the PGPE, we investigated the melting of a vortex lattice by perturbing the ground state of the system. Future work will focus on using the method to investigate dynamical phase transitions and vortex dynamics in rotating BECs.


\acknowledgments
We thank Tom Bland, Nick Parker, and Toby Wood for helpful discussions.  We would also like to thank A. L. Fetter who pointed out Ref.~\cite{Cozzini2006} to us.  R.D. thanks the Engineering and Physical Sciences Research Council of the UK (Grant No. EP/N509528/1) for support. This research made use of the Rocket High Performance Computing service at Newcastle University.


\appendix 
\section{The One-Body Hamiltonian}
\subsection{Normalisation and Orthonormality of the Eigenfunction}
\label{app_normalisation}
In this section we calculate the normalisation factor $A_n$ of the the wavefunction given in Eqn. \eqref{eqn_eigenfunction},
\begin{eqnarray}
\phi_{n,k} &=& A_n \sum_{p=-\infty}^\infty \chi_n \left[ \Gamma a \left( \frac{k}{N} + p \right) - \Gamma x \right] \exp \left[ i \Gamma^2 a \left( \frac{k}{N} + p  \right) y \right], \nonumber 
\end{eqnarray}
with
\begin{equation*}
\chi_n (x) = \frac{1}{\sqrt{2^n n! \sqrt{\pi}}} H_n (x) \exp \left( - \frac{1}{2} x^2 \right) ,  
\end{equation*}
such that 
\begin{equation}
ab  = \int_0^a \int_0^b \phi_{m,j}^* \phi_{n,k} \ dy \, dx . 
\label{eqn_normalization-condition}
\end{equation}
We need to assume that the summation converges in such a way that we may interchange the order of summation and integration. Then, the $y$--integral is
\begin{eqnarray}
I_y &=& \int_0^b \exp \left[ - i \Gamma^2 a \left( \frac{j}{N} + q \right) y \right] \exp \left[ i \Gamma^2 a \left( \frac{k}{N} + p \right) y \right] \ dy \nonumber \\
&=& \int_0^b \exp \left[ i \Gamma^2 a y \left( \frac{k-j}{N} + p-q \right) \right] \ dy.
\end{eqnarray}
We make the substitution $2 \pi y = b \tilde{y}$ so that for $\tilde{y} \in [0, 2\pi )$ we have
\begin{eqnarray}
I_y &=& \frac{b}{2\pi} \int_0^{2 \pi} d\tilde{y} \ \exp \left[ i \Gamma^2 \frac{ab}{2 \pi} \tilde{y} \left( \frac{k-j}{N} + p-q \right) \right] \nonumber \\
&=& \frac{b}{2\pi} \int_0^{2 \pi} d\tilde{y} \ \exp \left[ i \tilde{y} \left( k-j +Np - Nq \right) \right] .
\end{eqnarray}
We are now in a position where, since $j,k,p,q,N \in \mathbb{Z},$  we can apply the identity
\begin{equation}
\int_0^{2 \pi} e^{inx} = \begin{cases}
2 \pi, \qquad n=0 \\ 0, \qquad \text{otherwise} .
\end{cases}
\end{equation}
In order that $I_y$ doesn't vanish, we have the requirement $\left[ k-j + N \left( p-q \right) \right] = 0.$ This condition is separable , however, as $k,j \in \{0,1,\dots,N-1\},$ thus
\begin{equation}
I_y = b \, \delta_{j,k} \, \delta_{p,q}.
\end{equation}
The result for $I_y$ now reduces Eqn. \eqref{eqn_normalization-condition} to
\begin{eqnarray}
 & \, & \int_0^a \int_0^b \phi^*_{m,j} \phi_{n,k} \ dx \, dy =  A_m^* A_n b \times \nonumber \\
 & \, & \sum_{p=-\infty}^\infty  \int_0^a \chi^*_m \left[ \Gamma a \left( \frac{k}{N} + p \right) - \Gamma x \right] \chi_n \left[ \Gamma a \left( \frac{k}{N} + p \right) - \Gamma x \right] \ dx . \nonumber \\
\end{eqnarray}
We note that the Hermite functions, $\chi_m$ are real, and that the summation over $p,$ imposed to provide the periodic boundary conditions of the solution, essentially transforms the integral into an infinite domain, such that 
\begin{equation*}
\int_0^a \int_0^b \phi^*_{m,j} \phi_{n,k}  \, dx  dy  = A_m^* A_n \frac{b}{\Gamma} \int_{- \infty}^{\infty} \chi_m \left( \tilde{x} \right) \chi_n \left( \tilde{x} \right) \ d \tilde{x} .
\end{equation*}
The Hermite polynomials, $H_n (x)$ are orthogonal over $( - \infty, \infty)$ with respect to the weight function $e^{-x^2},$ so the Hermite functions $\chi_n(x),$ defined in Eqn. \eqref{eqn_hermite_function}, are orthonormal over this interval. This leaves 
\begin{equation}
ab  = A^*_m A_n \frac{b}{\Gamma} \, \delta_{m,n} 
\end{equation} 
so 
\begin{eqnarray}
\phi_{n,k} = \sqrt{a \Gamma}  \sum_{p=-\infty}^\infty \chi_n \left[ \Gamma a \left( \frac{k}{N} + p \right) - \Gamma x \right] \exp \left[ i \Gamma^2 a \left( \frac{k}{N} + p  \right) y \right]. \nonumber \\
\end{eqnarray}


\subsection{Quasi-Periodicity of the Eigenfunction}
\label{app_periodicity-of-eigenfunctions}
We can also show that $\phi_{n,k}$ obeys the quasi-periodic boundary conditions given in Eqns. \eqref{eqn_xbc} -- \eqref{eqn_ybc}. The $y$--direction is trivial, as taking $y \to y+b$ gives
\begin{eqnarray}
& \ & \phi_{n,k}(x,y+b) \nonumber \\
& = & A_n \sum_{p=-\infty}^\infty \chi_n \left[ \Gamma a \left( \frac{k}{N} + p \right) - \Gamma x \right] e^{ i \Gamma^2 a \left( \frac{k}{N} + p  \right) y } e^{i \Gamma^2 \frac{ab}{N} \left( k + Np  \right)} \nonumber \\
&=& A_n \sum_{p=-\infty}^\infty \chi_n \left[ \Gamma a \left( \frac{k}{N} + p \right) - \Gamma x \right] e^{ i \Gamma^2 a \left( \frac{k}{N} + p  \right) y } e^{2 \pi i \left( k + Np  \right)},
\end{eqnarray}
which is in agreement with Eqn. \eqref{eqn_ybc}. On setting$x \to x+a$ we get
\begin{eqnarray}
& \ & \phi_{n,k}(x,y+b) \nonumber \\
& = & A_n \sum_{p=-\infty}^\infty \chi_n \left[ \Gamma a \left( \frac{k}{N} + p \right) - \Gamma x - \Gamma a  \right] \exp \left[ i \Gamma^2 a \left( \frac{k}{N} + p  \right) y \right]  \nonumber \\
& = & A_n \sum_{p'=-\infty}^\infty \chi_n \left[ \Gamma a \left( \frac{k}{N} + p' \right) - \Gamma x \right] \exp \left[ i \Gamma^2 a \left( \frac{k}{N} + p'  \right) y \right]  e^{ i \Gamma^2 a y}  \nonumber \\
&=& \phi_{n,k}(x,y) \exp \left( i \frac{2 \pi N y}{b} \right) , 
\end{eqnarray} 
where $p' = p-1$. Taking the principal value of the argument of this, we recover
\begin{equation*}
\Arg \left[ \phi_{n,k} \left( x+a, y \right) \right] = \Arg \left[ \phi_{n,k} \left(x,y \right) \right] + \frac{2 \pi y}{b},
\end{equation*}

which is Eqn. \eqref{eqn_xbc}.



\end{document}